\documentclass[onecolumn,aps,prl]{revtex4}

\usepackage{graphicx}
[1999/03/29 PSNFSS v.7.2
Times font as default roman
: S Rahtz]

\begin{document}
\renewcommand{\thefootnote}{\fnsymbol{footnote}}
\sloppy
\newcommand{\rp}{\right)}
\newcommand{\lp}{\left(}
\newcommand \be  {\begin{equation}}
\newcommand \ba {\begin{eqnarray}}
\newcommand \bas {\begin{eqnarray*}}
\newcommand \ee  {\end{equation}}
\newcommand \ea {\end{eqnarray}}
\newcommand \eas {\end{eqnarray*}}

\title{Communication impacting financial markets}
\thispagestyle{empty}

\author{J\o rgen Vitting Andersen$^1$, Ioannis Vrontos$^2$, Petros Dellaportas$^2$ 
and  Serge Galam$^3$
\vspace{0.5cm}}
\affiliation{$^1$
CNRS, Centre d\'{}Economie de la Sorbonne,  
Universit\'e Paris 1 Panth\'eon-Sorbonne, \\
Maison des Sciences Economiques,
106-112 Boulevard de l’H\^opital 75647 Paris Cedex 13, France \\}
\email{Jorgen-Vitting.Andersen@univ_paris1.fr}
\affiliation{$^2$Department of Statistics \\
Athens University of Economics and Business \\
Patission 76,104 34 Athens, Greece\\}
\affiliation{$^3$Cevipof - Center for Political Reseach, \\
Sciences Po and CNRS,\\
98 rue de l'Universit\'e, 75007 Paris, France\\}

\begin{abstract}

Since the attribution of the Nobel prize in 2002 to Kahneman for prospect theory, behavioral finance has become an increasingly important subfield of finance. However the main parts of behavioral finance, prospect theory included, understand financial markets through {\em individual} investment behavior. Behavioral finance thereby ignores any interaction between participants.

We introduce a socio-financial\cite{AN} model that studies the  impact of communication  on the pricing  in financial markets. 
Considering the  simplest possible case where each market participant  has either  a positive  (bullish) or negative (bearish) sentiment with  respect to  the market,  we model the evolution of the sentiment in the population due to communication  in subgroups of different sizes. Nonlinear feedback effects between the market performance and changes in sentiments are taking into account by assuming  that the market performance is dependent on changes in sentiments (e.g.  a large sudden positive change in bullishness would lead to more buying).  The market performance in turn has an impact  on the sentiment through  the transition  probabilities to change an opinion in a group of a given size. The idea is that  if for example the market has observed a recent downturn, it will be easier for even a bearish minority to convince  a  bullish majority  to change  opinion compared to the case where the  meeting  takes place in a bullish  upturn  of the market.

Within the framework of our proposed model, financial markets stylized facts such as volatility clustering  and extreme events may be perceived as arising due to abrupt sentiment changes via ongoing communication of the market participants.  
The  model introduces a new volatility measure which is apt of capturing volatility 
clustering and from maximum likelihood analysis we are able to  apply the  model to real  data and  give additional long term insight into where a  market is heading. 
\end{abstract}

\maketitle

\vspace{1cm}
\noindent

\section{Introduction}
The quote ``In the short run, the market is a voting machine, but in the long run 
it is a weighing machine'' is attributed to Benjamin Graham\cite{Graham}. 
Graham himself used the quote to argue that investors should use the so-called fundamental value investment approach and concentrate on analyzing accurately the worth of a given financial asset. That is, he suggested ignoring the short run
``voting machine'' part, and instead concentrate on the long run, where somehow the 
``weighing machine''of the market would ensure to end up with a pricing of the true worth of an asset.
The interesting part of Graham's quote is the allusion to the decision making of people and its impact on the markets. Graham however does not refer to how 
the decision making process actually takes places, and this will be our focus in the following.

In situations of uncertainty people often consult others to get more information 
and thereby a better understanding of the situation. This is in particular true 
with respect to financial markets, where market participants consult the media or 
other colleagues to get an idea of the origin behind price movements, or to assess 
which impact a given information could have on the markets. For example, copying the behavior of others is one of the 
predominant mechanisms in decisions to purchase\cite{SHW}. Social households, i.e.  those that interact with their neighbors, 
have also been shown to have an impact in the level of stock market investments. In \cite{HKS} it was shown how the stock-market participation rates were higher in places that had a  higher sociability. 
However the sole knowledge of the role of imitation is not sufficient to describe in an operative frame the price dynamics of financial markets since indeed the rules governing opinion dynamics and group decisions still needs to be understood and identified in more detail.
In \cite{SP} interpersonal communication were found to 
be extremely important in investors decisions. Questionnaire surveys of institutional and individual investors were found to reveal 
a strong influence by word-of-mouth communications. 
In \cite{Arnswald} it was shown that among fund managers in Germany information exchange with other financial 
and industry experts was the second most important factor influencing their investment decisions, complemented by
conversations with their colleagues and reports from media.
More recently impact of large scale human network  structes and its  
flow of information were shown to have an  impact on the asset pricing\cite{OW}.
Different modeling effort was presented in\cite{Bornholdt} where a spin model was introduced to take into account local interaction between different market participants to explain the formation of  bubbles in financial markets. A follow up study was presented  in\cite{KBF} showing aperiodic switching between bull markets and bear markets due to the local heterogeneous interaction of traders. A different view  of  interaction between  market participants was presented in an ``ant recruitment'' model of 
herding/epidemics in\cite{Kirman}. In similar vein models of social opinion dynamics and agent based models of investor sentiments was presented in \cite{Lux1, Lux2, Lux3, LM, Topol, AM}. It should be  noted however that  only the work of \cite{Lux1, Lux3} keep the price and sentiments as two distinguished variables which we find important since sentiment (as well as price) can be obtained empirically.
In the following we will introduce a socio-financial\cite{AN} model that in a quantitative manner 
does exactly this. 

\section{Analysis}

Consider a population of market participants, shown schematically as circles in Fig.~1A. We will proceed as in the so-called Galam model of opinion  formation\cite{G1,G2,BGG} 
and for simplicity imagine that people have just two different opinions on 
the market, which we can characterize as either "`bullish"' (black circles) or "bearish" (white circles). Letting $B(t)$ denote the proportion of bullishness in a population at time $t$, the proportion of bearishness is then $1-B(t)$.

\begin{figure}
\includegraphics[width=14cm]{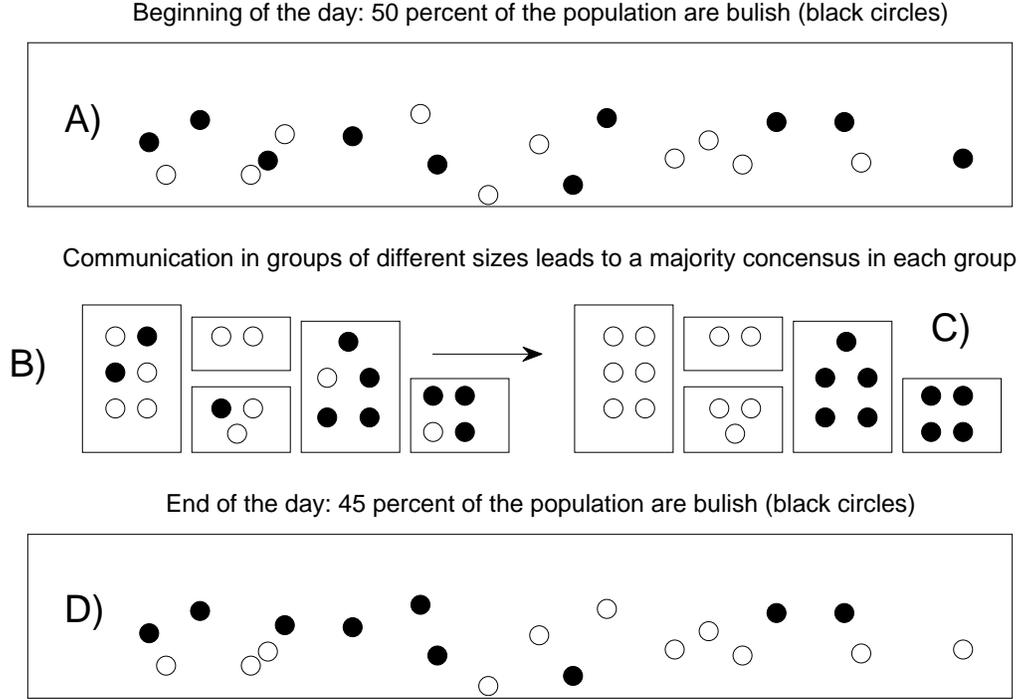}
\caption{\protect\label{Fig1}
Changing the ``bullishness'' in a population via communications in subgroups. {\bf a):}  At 
the beginning of a given day $t$ a certain percentage $B(t)$ of bullishness. {\bf b):} During the 
day communication takes place in random subgroups of different sizes. {\bf c):} illustrates the 
extreme case of complete polarization $m_{k,j} = \pm 1$ created by a majority rule in 
opinion. In general $m_{k,j} \simeq j/k$ corresponds to the neutral case where in average 
the opinion remains unchanged within a subgroup of size $k$. {\bf d):} due to the communication in 
different subgroups the ``bullishness'' at the end of the day is different from the beginning of the day.
}
\end{figure}

Figure~1A  
represents the opinions of the participants at the beginning of a given day. During 
the day people meet in random subgroups of different sizes, as illustrated by the different boxes in Fig.~1B, to update their view of 
the market. Take, for example, the leftmost box in Fig.~1B with six persons, two bullish, four bearish, 
 who we can imagine are sitting around a table, or having a conference call, discussing 
the latest market developments. The outcome of the discussions for the different groups are illustrated in 
Fig.~1C. For simplicity we have illustrated the case where a majority opinion in a given 
subgroup manages to polarize the opinion of the group by changing the opinion of 
those who had an opinion belonging to the minority. If we take the afore mentioned group of six 
persons we can see that after discussing, because of the majority polarizing rule,  they have all become bearish. More realistically, we will 
in the following instead 
assume that is a certain {\em probability} for a majority opinion to prevail, and that  
even under certain conditions a minority could persuade a part of the majority 
to change their opinion.

For a given group of size $k$ with $j$
agents having a bullish opinion and $k - j$ a bearish opinion, we let $m_{k,j}$ denote the transition
probability for all ($k$) members to adopt the bullish opinion as a result of their
meeting. After one update taking into account communications in all groups of size $k$ with $j$ bullish agents, 
the new probability of finding an agent with a bullish
view in the population can therefore be written:
\be
\label{B_kj}
B(t+1)  =  m_{k,j} (t) C_j^k B(t)^j [1 - B(t)]^{k-j}
\ee
where
\be 
\label{C_kj}
C^k_j \equiv {k! \over j! (k-j)! }
\ee
are the binomial coefficients. Notice that the transition probabilities $m_{k,j}$ depend on time, 
since we assume that they change as the market performance changes (this point will be explained further below).

Taking the sum over different groups of different sizes and different composition of bullishness within each group(see Fig.~1B) one obtains a general term, $B(t+1)$, for the bullishness in a  population at time $t+1$ due to the outcome of meetings of groups with different sizes and different composition of bullishness: 
\ba
\label{B_t}
B(t+1) &  = &   \sum_{k=1}^{L} a_k \sum_{j=0}^k m_{k,j} (t) C^k_j B(t)^j [1-B(t)]^{k-j} \\
\sum_{k=1}^L a_k = 1 & ;  & a_k \equiv {1 \over L}
\ea
With $L$ denoting the size of the largest group and $a_k$ denoting the weight of the group of size $k$.
The link between communication and its impact on the markets can then be taken
into account by assuming that the price return $r(t)$ changes whenever there
is a change in the bullishness. The idea is that the bullishness itself is not
the relevant factor determining how prices will change. Those feeling bullish
would naturally already hold long positions on the market. Rather, when
people change their opinion, say becoming more negative about the market,
or less bullish, this will increase their tendency to sell. 
 The fact that the absolute sentiment can act as a contrarian indictor (and the change in sentiment as an indicator) for future  market returns is well known among practitioners, see e.g. \cite{Hulbert} which present a sentiment index used for contrarian predictions of future market returns.

Assuming the return to
be proportional to the percentage change in bullishness, $RB(t)$, as well as economic
news, $\eta (t)$, the return $r(t)$ is given by:
\be
\label{r_t}
r(t)  =  { RB(t) \over \lambda} + \eta (t) , \ \ \lambda > 0 \\
\ee
with $RB(t) = {B(t)-B(t-1) \over B(t-1)}$ the change or ``return'' of the bullishness.  The variable $\eta(t) = r(t) - {RB(t) \over \lambda}$ is assumed to be Gaussian distributed with a mean $\mu \equiv 0$ and a standard deviation that varies as a function of time 
depending on changes in sentiment.  We will assume that the market will react to fundamental economic news represented by $\eta$ but 
that the amplitude of the reaction depends on changes in the sentiment $RB(t)$:

\ba
\label{sigma}
\sigma (t)  & =  & \sigma_0 \exp{({|RB(t)| \over \beta })} , \ \ \sigma_0 > 0, \ \ \beta > 0
\ea

The influence of the financial market on decision-making can now be included 
in a natural way by letting the strength of persuasion depend on how the market has 
performed since the last meeting of the market participants. The idea is that, if for 
example the market had a dramatic downturn at the close yesterday, then in meetings 
the next morning, those with a bearish view will be more likely to convince even a 
bullish majority of their point of view. In the formal description below, this is taken 
into account by letting the transition probabilities for a change of opinion, i.e., the 
probabilities of transitions like Fig.~1.B $\rightarrow$ Fig.~1.C, depend on the market return over the 
last period:

\ba
\label{m_t}
m_{k,j} (t) & = & m_{k,j} (t-1) \exp{({r(t) \over \alpha })}; m_{k,j} (t=0) \equiv j/k , \ \ \alpha > 0
\ea
where $\alpha$ defines the scale for which a given return $r(t)$ impacts the transition probabilities. The 
condition $m_{k,j} (t=0) \equiv j/k$ describes the initially unbiased case where in average no market participant changes opinion.

\section{Results}

As a first demonstration of the properties of the model, Eqs. (\ref{B_t}-\ref{m_t}),  
Figure~2a shows an illustration of the link between the bullishness of the market 
participants obtained through communication (thin solid line) and market prices 
(thick solid line). One observes a clear almost 
continuous decrease in bullishness, whereas the market itself first decreases but then 
regains what it lost at the end of the time period,  essentially ending up unchanged compared to its initial level. 
This illustrates the competition between persistence through memory effects in the bullishness (1-7) and the randomness 
of the return via the $\eta (t)$-term in (5). 
It should be noted that $B=0,1$ act as repulsive fix points: if $B$ converges towards these two extremes the derivative of $B$ goes to zero and one can neglect the sentiment part in the pricing formula (5).  In that case any fundamental negative news ($\eta < 0$) will not further impact the sentiment in a bearish population (say $B \approx. 0$) whereas  any positive fundamental news will lead to a change towards more positive sentiments via (7), (3).
The decaying rate of the return of the bullishness over the last part of 
the time period and the simultaneous rise  of the prices  
illustrates indeed the complex and highly nonlinear relationship between returns and sentiments. 
The model is able to reproduce the most important of the so-called 
stylized facts seen in real financial market data\cite{Cont}. Figure~2b illustrates clustering of the volatility of the return 
of the time series in Fig.~2a. Furthermore the returns show  
fat-tailed behavior in the returns (Fig.~2c), with fat tail exponents similar to what is found in real markets\cite{Stanley}. 
Finally let us also mention that there are no arbitrage possibilities within the framework of the model seen by the zero autocorrelations of 
returns (thin line, Fig.~2d), whereas long memory effects are seen in the volatility, seen via the slowly decaying autocorrelations
 of volatility (thick line, Fig.~2d). 
It should be noted that fixing the other parameters we find the pairwise $(L=2)$ case  to show similar behavior as the general $L$ case (we have only studied $L<20$). A similar conclusion was found using different distributions of $a_k$, which seems to indicate a robustness of contagion effects on those two parameters somewhat similar to what was reported in \cite{AM}.

\begin{figure}[h]
\includegraphics[width=14cm]{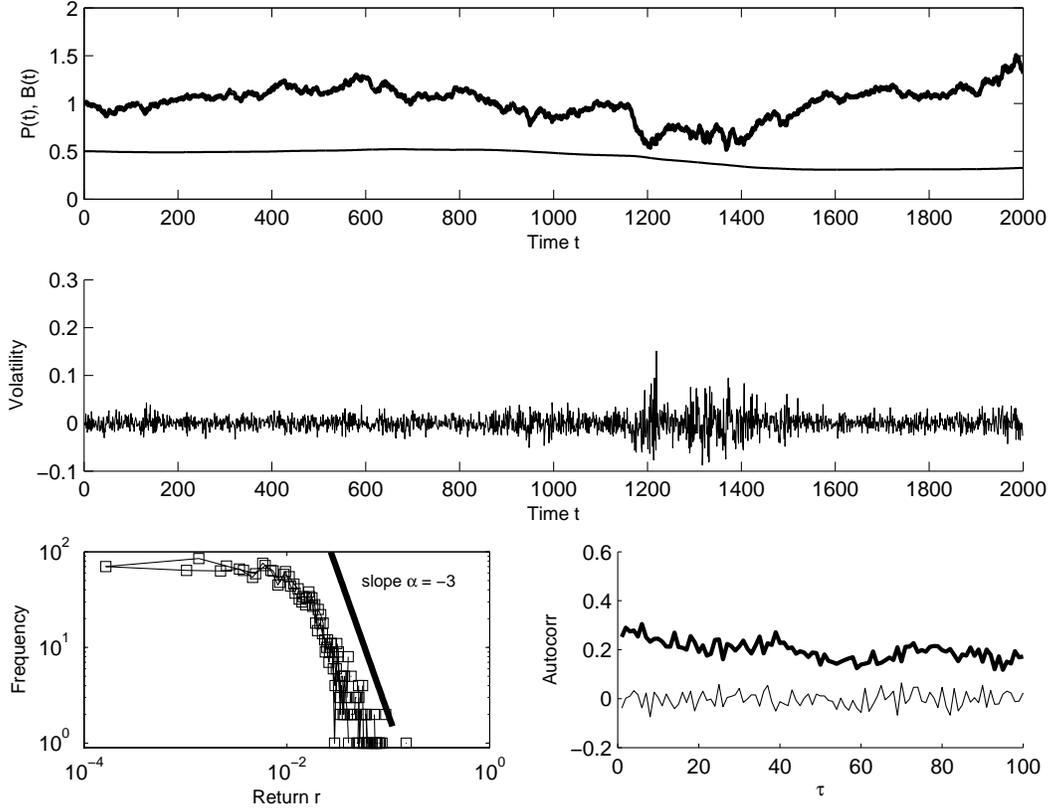}
\caption{\protect\label{Fig2}
Reproducing the ``stylized facts''. (a) Example of how a change in bullishness $B(t)$ in a population (thin line) 
can have an impact on prices $P(t)$ (thick solid line). (b) Volatility clustering as a function of time. (c) 
''Fat tailed'' returns. (d) No arbitrage possibilities, i.e.  zero autocorrelations of returns (thin solid line), 
and  long time memory effects in volatility, i.e. nonzero autocorrelations of volatility versus time (thick solid 
line). Parameter values used: $\lambda=1.1$, $\sigma_0 = 0.01$, $\beta = 0.001$, $L = 5$.
}
\end{figure}

Using inference based on maximum likelihood (see  appendix)  
we first 
conducted a series of
simulations in order to ensure indeed to be able to retrieve the  parameter  values used to generate the {\em simulated} data\cite{work_in_progress}. 
Next, we examined an empirical application of the proposed
socio-financial model to the FTSE-20 Athens stock exchange index. The idea is to consider a very 
volatile market to study abrupt and large changes  in  market performance and map the corresponding 
evolution in sentiments. The data
consists of $1247$ daily prices over the 7/11/2008-6/11/2013 period. 

The results showed evidence of convergence of the maximization algorithm
since the estimates found by using different starting values are very
similar, and the value of the gradient of the log-likelihood evaluated at
the parameter estimates is near zero. 

We then estimated the proportion of bullishness across time and the
time-varying conditional volatilities. In Figure~3, we present the FTSE-20
prices $P(t)$, the corresponding returns $r(t)$, the
estimated bullishness proportions $\widehat{B}(t)$ and the estimated
conditional volatilities $\widehat{\sigma }(t)$, which are based on the
parameter estimates of the socio-financial model. Comparing the FTSE-20 price 
evolution Figure~3(a) and the estimated bullishness proportions Figure~3c, one observes a 
steady decline in the bullishness proportions even after a relatively long (two years: from 
mid-2011 to mid-2013)  and stable price evolution. We consequently suggest to consider this 
as indicating a `bad signal'' for the following FTSE-20 price levels.
Finally, comparing the observed volatility of the FTSE-20, Figure~3(b), and the estimated volatility from 
the model, Figure~3(d), one notices  that periods of high volatility of the market is indeed detected by 
 volatility estimates of the socio-financial model. Therefore our 
proposed model can be seen as introducing an alternative measure to capture volatility clustering phenomena   
observed in return series of financial market data.

Looking at figure 3.1 one can conclude that for the given parameter choices the   main part of the $RB(t)$ choice is not  on the return but on the volatility. However different choices of $\gamma, \sigma_0$ and $\beta$ could make a stronger impact on the sentiments changes directly on the returns on a shorter time scale. As such the model can be imagined to describe a slow (say monthly/yearly) impact of changes of sentiments on the returns due to possible real economic factors, and a faster (i.e. on a higher frequency, say daily) change in the susceptibility of reaction to economic news depending on  changes in sentiments.  The ratio $\lambda/\beta$ therefore controls the slow versus fast time scale of the  price formation.

\begin{figure}[h]
\includegraphics[width=14cm]{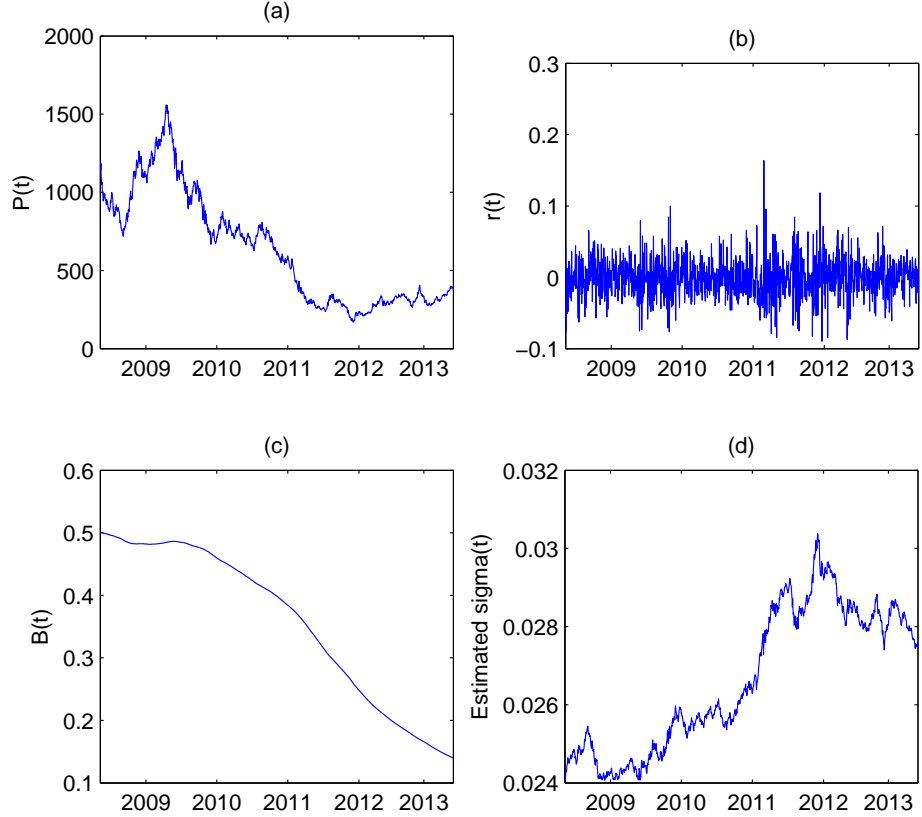}
\caption{\protect\label{Fig4}
FTSE-20
Athens stock index prices and returns, as well as the corresponding
estimated conditional volatilities and bullishness proportions using optimal parameters estimated 
via the  maximum likelihood method. (a) FTSE-20
price $P(t)$, (b) FTSE-20 returns $r(t)$,  (c) Estimated bullishness
proportions $\widehat{B}(t)$. (d) Estimated conditional
volatilities $\widehat{\protect\sigma }(t)$.
}
\end{figure}

We emphasize here that the econometric analysis presented above is just an illustration of the empirical properties of the proposed model and not a thorough statistical analysis that would require careful assessment of the assumptions adopted.  In a preliminary analysis we have found that the density of $\eta(t)$ could be replaced by a student-t density producing better fit to the particular data; this is not surprising as it is well-known that in the popular GARCH models a student-t  error distributional assumption is preferred to normal.

\section{Discussion}

The main parts of traditional finance as well as behavioral finance, prospect theory\cite{KT} included, understand pricing in financial markets through {\em individual} investment behavior, thereby ignoring any interaction between market participants. 
We have on the contrary proposed  a socio-financial model that focus on interaction between market participants via communication. 
Opinion  formation of  the market participants is modeled by communication that takes  place in  subgroups of different  sizes.  Nonlinear feedback effects were introduced to describe how market performance can change sentiments and vice versa.
Our socio-financial model was shown to be able to reproduce the main stylized facts of financial market data.
By  estimating the model parameters via maximum likelihood method we are able to study extreme behavior of markets swings and sentiments  illustrated by a case of  the FTSE-20 Athens stock index over the last five years. Recent price behavior of the FTSE-20 index seems to indicate a reversal of the bearish trend whereas the sentiment measure obtained from our method seems to indicate that the market still has room for further downturns.
Finally our model introduces a new volatility measure apt of capturing volatility 
clustering seen in the real financial market data

\section{Appendix}

In the following we present an inferential method adopted to estimate the
parameters of the model (3-7) based on the maximization of the conditional
likelihood. Below we describe the calculation of the likelihood function for
the model for a sample of $T$ observations $r=\left(
r(1),r(2),...,r(T)\right) $ under the assumption of a Normal distribution
for the error process $\eta (t)$.

Consider the probability distribution of $r(1)$, the first observation in
the sample. Since $\eta (t)$ is assumed Gaussian, the density of the first
observation, conditional on $B(0)$ and $m_{k,j}(0)=j/k$ (corresponding to
the neutral case), takes the form%
\[
f\left[ r(1)|B(0),m_{k,j}(0),\mathbf{\theta }\right] =\frac{1}{\sqrt{2\pi }%
\sigma (1)}\exp \left\{ -\frac{1}{2\sigma (1)^{2}}\left[ r(1)-\frac{1}{%
\lambda }RB(1)\right] ^{2}\right\} ,
\]%
where $\mathbf{\theta =(}\lambda $, $\sigma _{0}$, $\beta $, $\alpha
)^{\prime }$ denotes the parameter vector to be estimated. Next,
conditioning on $r(1)$, the density of the second observation $r(2)$ is%
\[
f\left[ r(2)|r(1),B(0),m_{k,j}(0),\mathbf{\theta }\right] =\frac{1}{\sqrt{%
2\pi }\sigma (2)}\exp \left\{ -\frac{1}{2\sigma (2)^{2}}\left[ r(2)-\frac{1}{%
\lambda }RB(2)\right] ^{2}\right\} .
\]%
Proceeding in this fashion, the conditional density of the $t-th$
observation can be calculated as%
\[
f\left[ r(t)|r(t-1),...,r(1),B(0),m_{k,j}(0),\mathbf{\theta }\right] =\frac{1%
}{\sqrt{2\pi }\sigma (t)}\exp \left\{ -\frac{1}{2\sigma (t)^{2}}\left[ r(t)-%
\frac{1}{\lambda }RB(t)\right] ^{2}\right\} .
\]%
Therefore, the likelihood of the complete sample can be written as%
\[
f\left[ r|B(0),m_{k,j}(0),\mathbf{\theta }\right] =\left( 2\pi \right) ^{-%
\frac{T}{2}}\prod\limits_{t=1}^{T}\sigma (t)^{-1}\cdot \exp \left\{ -\frac{1%
}{2}\sum\limits_{t=1}^{T}\frac{1}{\sigma (t)^{2}}\left[ r(t)-\frac{1}{%
\lambda }RB(t)\right] ^{2}\right\} .
\]%
The log-likelihood function, denoted $L_{T}\left( r|\mathbf{\theta }\right) $%
, can be written as%
\[
L_{T}\left( r|\mathbf{\theta }\right) =-\frac{T}{2}\ln \left( 2\pi \right)
-\sum\limits_{t=1}^{T}\ln \sigma (t)-\frac{1}{2}\sum\limits_{t=1}^{T}\frac{1%
}{\sigma (t)^{2}}\left[ r(t)-\frac{1}{\lambda }RB(t)\right] ^{2}.
\]%
Clearly, the value of $\mathbf{\theta }$ that maximizes the conditional
likelihood is identical to the value that maximizes the conditional
log-likelihood.

The population dynamics enters in $L_{T}(r | \theta)$ via $RB(t)$  and $\sigma (t)$ through  (1) and (6).

\section{Acknowledgments}
The research leading to these results has received funding from the European Union Seventh Framework Programme (FP7-SSH/2007-2013) under grant agreement n° 320270 – SYRTO.

\end{document}